\documentclass[aps,twocolumn,floatfix, bibnotes]{revtex4}

\usepackage{graphicx}%
\usepackage{dcolumn}
\usepackage{color}
\usepackage{amsmath}
\usepackage{here}

\begin{document}
\begin{sloppy}

\newcommand{\be}{\begin{equation}}
\newcommand{\ee}{\end{equation}}
\newcommand{\bea}{\begin{eqnarray}}
\newcommand{\eea}{\end{eqnarray}}
\newcommand\bibtranslation[1]{English translation: {#1}}
\newcommand\bibfollowup[1]{{#1}}

\newcommand\pictc[5]{\begin{figure}
                       \centerline{
                       \includegraphics[width=#1\columnwidth]{#3}}
                  \protect\caption{\protect\label{fig:#4} #5}
                    \end{figure}            }
\newcommand\pict[4][1.]{\pictc{#1}{!tb}{#2}{#3}{#4}}
\newcommand\rpict[1]{\ref{fig:#1}}

\newcommand\leqt[1]{\protect\label{eq:#1}}
\newcommand\reqtn[1]{\ref{eq:#1}}
\newcommand\reqt[1]{(\reqtn{#1})}

\newcounter{Fig}
\newcommand\pictFig[1]{\pagebreak \centerline{
                   \includegraphics[width=\columnwidth]{#1}}
                   \vspace*{2cm}
                   \centerline{Fig. \protect\addtocounter{Fig}{1}\theFig.}}

\title{Backward Tamm states in left-handed metamaterials}

\author{Abdolrahman Namdar$^{1,2}$, Ilya V. Shadrivov$^1$, and Yuri S. Kivshar$^1$}

\affiliation{ $^1$Nonlinear Physics Center, Research School of
Physical Sciences and Engineering, Australian National University,
Canberra ACT 0200,
Australia\\
$^2$Physics Department, Azarbaijan University of Tarbiat
Moallem, Tabriz, Iran}

\begin{abstract}
We study the electromagnetic surface waves localized at an interface
separating a one-dimensional photonic crystal and left-handed
metamaterial, the so-called surface Tamm states. We demonstrate that
the metamaterial allows for a flexible control of the dispersion
properties of surface states, and can support the Tamm states with a
backward energy flow and a vortex-like structure.
\end{abstract}


\maketitle

The study of unusual properties of {\em left-handed metamaterials}
is a rapidly developing field of physics. Left-handed
metamaterial (LHM)  is characterized by simultaneously negative
effective dielectric permittivity and negative effective magnetic
permeability, that gives rise to a variety of unusual properties of
electromagnetic waves. Such materials have first been suggested
theoretically by Veselago~\cite{Veselago:1967-517:UFN} almost 40
years ago, but they have been realized experimentally only
a few years ago~\cite{Smith:2000-4184:PRL}. Later, it was shown that they
possess many extraordinary wave-guiding properties~\cite{Shadrivov:2003-57602:PRE,Ruppin:2001-1811:JPCM,Ruppin:2000-61:PLA,Shadrivov:2004-16617:PRE}. In this Letter we demonstrate another example of unusual properties of LHM and study electromagnetic surface waves guided by an
interface between LHM and a one-dimensional photonic crystal, the
so-called {\em surface Tamm states}~\cite{Tamm:1932-733:PZS}. We
demonstrate that the presense of a metamaterial allows for a flexible control of
the dispersion properties of surface states, and the interface can support the
Tamm states with a backward energy flow and a vortex-like
structure.

Surface modes are a special type of waves localized at an interface
between two different media. In periodic systems, staggered modes
localized at surfaces are known as Tamm
states~\cite{Tamm:1932-733:PZS,Shockley:1939-317:PREV}, first found
in solid-state physics as localized electronic states at the edge of
a truncated periodic potential. Surface states have been studied in
many different fields of physics, including
optics~\cite{Yeh:1977-423:JOS, Yeh:1978-104:APL}, where such waves
are confined to an interface between periodic and homogeneous
dielectric media, as well as nonlinear dynamics of discrete
chains~\cite{physica_d}. In optics, the periodic structures have to
be manufactured artificially in order to manipulate dispersion
properties of light in a similar way as the properties of electrons
are controlled in crystals. Such periodic dielectric structures are
known as photonic crystals. An analogy between solid-state physics and
optics suggests that surface electromagnetic waves should exist at
the interfaces of photonic crystals, and indeed they were predicted
theoretically~\cite{Yeh:1977-423:JOS, Yeh:1978-104:APL} and observed
experimentally~\cite{Robertson:1999-1800:APL}. Such Tamm states can
be very important for applications of photonic crystals, as they
allow for the enhanced coupling of the electromagnetic waves to and
from the photonic crystal
waveguides~\cite{Moreno:2004-121402:PRB,Morrison:2005-81110:APL}.

In this Letter, we study surface electromagnetic waves, or surface
Tamm states, guided by an interface separating homogeneous LHM and
one-dimensional photonic crystal. We assume that the terminating
layer (or a cap layer) of the periodic structure has the width
different from the width of other layers of the structure. We study
the effect of the width of this termination layer on surface states, and explore a possibility to control the dispersion properties of surface waves by
adjusting termination layer thickness. We find {\em novel types of
surface Tamm states} at the interface with the metamaterial which have a
backward energy flow and a vortex-like structure. We also compare
our results with the case when the LHM medium is replaced by a
conventional dielectric, which we refer to as right-handed material
(RHM). The surface states in these two cases we call left- and
right-handed Tamm states, respectively.

\pict[0.8]{fig01}{tamm_geom}{Geometry of the problem. In our calculations we take the following values: $d_1=1$cm, $d_2=1.65$cm,
$\varepsilon_1=4$, $\mu_1 = 1$, $\varepsilon_2=2.25$, $\mu_2 = 1$, $\varepsilon_0=-1$, and $\mu_0 = -1$.}

Geometry of our problem is sketched in Fig.~\rpict{tamm_geom}. We
consider the propagation of TE-polarized waves described by one
component of the electric field, $E=E_y$~\cite{note}, and governed by a
scalar Helmholtz-type equation. We look for stationary solutions
propagating along the interface with the characterestic dependence
$\sim \exp{\left[-i \omega(t - \beta x/c)\right]}$, where $\omega$
is an angular frequency, $\beta$ is the normalized wavenumber
component along the interface, and $c$ is the speed of light, and
present this equation in the form,
\be \leqt{Helm}
   \left[ \frac{d^2}{d z^2}
          + k_x^2
          + \frac{\omega^2}{c^2} \varepsilon(z) \mu(z)
          - \frac{1}{\mu(z)} \frac{d \mu}{d z}
            \frac{d}{d z} \right] E
   = 0,
\ee
where $k_x^2 = \omega^2 \beta^2/c^2$, both $\epsilon (z)$ and $\mu(z)$
characterize the transverse structure of the media. Surface modes
correspond to localized solutions with the field $E$ decaying
from the interface in both the directions. In a left-side
homogeneous medium ($z<-d_s$, see Fig.~\rpict{tamm_geom}), the fields
are decaying provided $\beta > \varepsilon_0 \mu_0$. In the
right-side periodic structure, the waves are the
Bloch modes,
\be
E(z) = \Psi(z)\exp{(i K_b z)},
\ee
where $K_b$ is the Bloch wave number, and $\Psi(z)$ is the Bloch
function which is periodic with the period of the photonic
structure (see details, e.g., in Ref.~\cite{Yeh:1988:OpticalWaves}).
In the periodic structure the waves will be decaying provided $K_b$ is complex; and this condition defines the spectral gaps of an infinite photonic crystal. For the calculation of the Bloch modes, we use the well-known transfer matrix
method~\cite{Yeh:1988:OpticalWaves}.

To find the Tamm states, we take solutions of Eq.~\reqt{Helm} in a
homogeneous medium and the Bloch modes in the periodic structure and
satisfy the conditions of continuity of the tangential
components of the electric and magnetic fields at the interface between homogeneous medium and periodic structure~\cite{Martorell:2005-0512272:ARX}. We summarize the dispersion properties of the Tamm states in the first and second spectral gaps on the plane of the free-space wavenumber $k = \omega/c$ versus
the propagation constant $\beta$ (see Fig.~\rpict{tamm_dispers}) for
different values of the cap layer thickness $d_c$. For comparison,
we also plot the dispersion of the corresponding Tamm states in the
structure, where the homogeneous medium is replaced by vacuum
(dashed).

\pict{fig02}{tamm_dispers}{Dispersion properties of the Tamm states in the first and second spectral gaps. Shaded: bands of the one-dimensional photonic crystal. Solid: dispersion of the Tamm states with LHM; dashed: dispersion of the surface waves in the problem where the metamaterial is replaced by vacuum. Corresponding values of the cap layer thickness are indicated next to the curves. Inset shows a blow-up region of small $\beta$. Points (a), (b), and (d) correspond to the mode profiles presented in
Figs.~\rpict{modes} (a,b,d).}

As mentioned above, the Tamm states exist in the gaps of the photonic
bandgap spectrum (unshaded regions in Fig.~\rpict{tamm_dispers}). For a
thin cap layer, $d_c = 0.01$cm, the left-handed Tamm state
approaches the lower edge of the first (lower) band gap, while
these right-handed Tamm state approaches the top edge. Another
important difference between the two cases, is that for the LHM
surface modes the slope of the dispersion curve becomes negative for
small $\beta$, and it remains positive for larger longitudinal
wavenumbers (see the inset in Fig.~\rpict{tamm_dispers}), while for
the conventional dielectric media, the dispersion curves are always with
a positive slope. The slope of the dispersion curve determines the
corresponding group velocity of the mode. The extended control over the group
velocity in the case of LHM bandgap structure is possible due to the
backward energy flow in metamaterials. Similar effects have been
already predicted for other types of the wave guiding
structures~\cite{Shadrivov:2003-57602:PRE,Shadrivov:2004-16617:PRE}.
Similarly, the left-handed Tamm surface wave has {\em a vortex-like energy flow pattern}.

\pict[0.9]{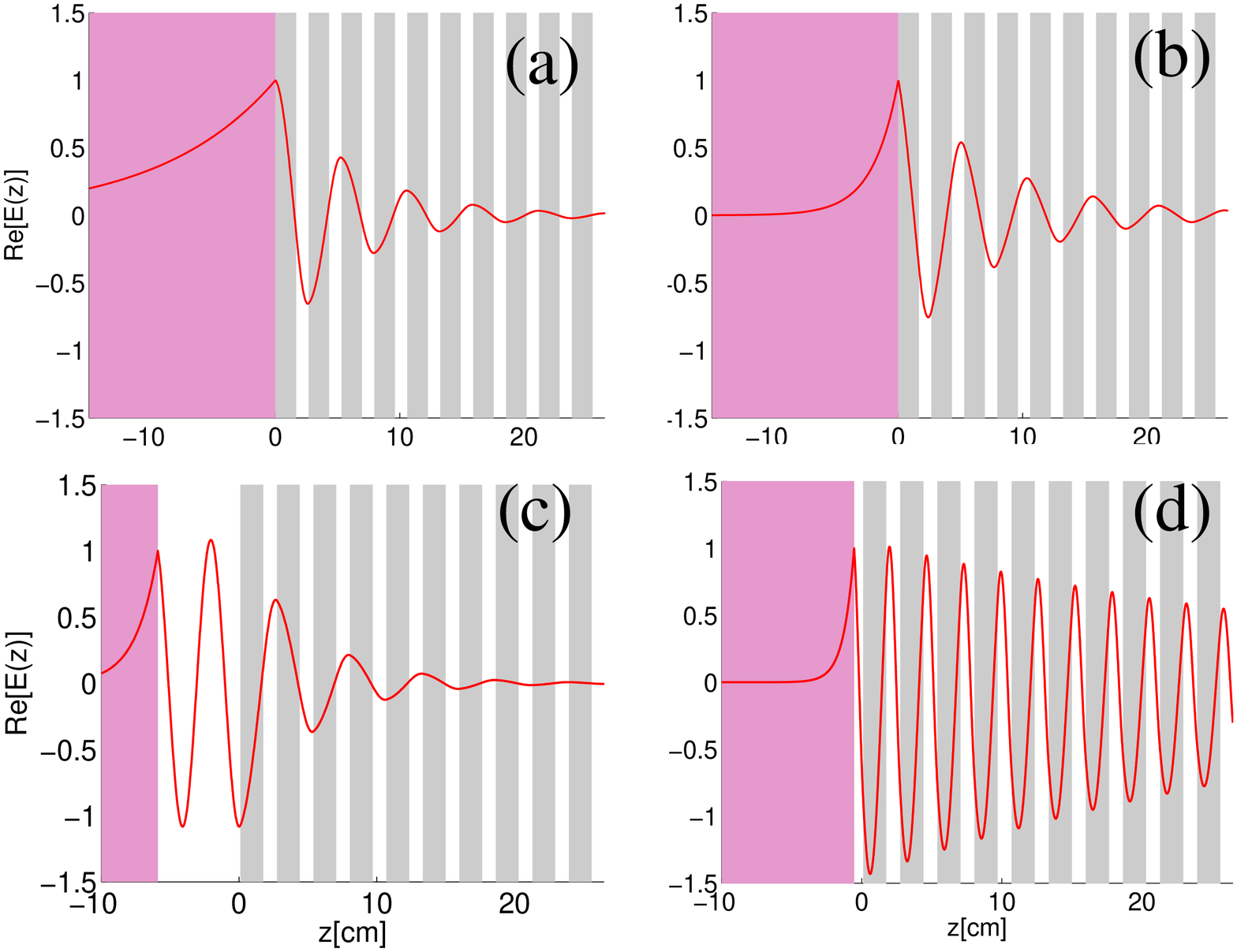}{modes}{Examples of the left-handed surface Tamm states. (a)
Backward LHM mode, $d_c=0.01$cm, $k=0.85$cm$^{-1}$, $\beta=1.008$;
(b) Forward LHM mode, $d_c=0.01$cm, $k=0.85$cm$^{-1}$,
$\beta=1.153$; (c) Guided mode with a thick cap layer, $d_c= 6$cm,
$k=0.9561$cm$^{-1}$, $\beta=1.2$; Surface mode in the second
bandgap,  $d_c= 0.65$cm, $k=2.0386$cm$^{-1}$, $\beta=1.2$. Modes
(a), (b), and (d) correspond to the points in
Fig.~\rpict{tamm_dispers}.}

As a result of different slope of the dispersion curve of the left-handed Tamm states, we observe the mode degeneracy, i.e., for the same frequency $\omega$
(or wavenumber $k$), there exist two modes with different value of $\beta$.
The mode with lower $\beta$ has a negative group velocity (with
respect to the propagation wavevector), while the other mode has a
positive group velocity. Such modes are termed as {\em backward} and
{\em forward}, respectively. In the forward wave, the direction of
the {\em total energy flow} coincides with the propagation direction,
while in the backward wave the energy flow is backwards with respect
to the wavevector. Physically, this difference can be explained by
looking at the transverse structure of these two modes. In Figs.~\rpict{modes}
(a,b) we plot the profiles of the two modes having the same frequency $k=\omega/c=0.85$cm$^{-1}$, with different longitudinal wavenumber $\beta$ . Corresponding points are shown in the inset in Fig.~\rpict{tamm_dispers}. For the mode (a), the energy flow in the metamaterial exceeds that in the periodic structure (slow decay of the field for $z<-d_s$ and fast decay into the periodic
structure), thus the total energy flow is backward. For the mode (b) we
have the opposite case, and the mode is forward. In the right-handed
Tamm state geometry, the energy flow in all parts of the wave are directed
along the wavenumber and the localized surface waves are always
forward. To demonstrate this, in Fig.~\rpict{power_flow} we plot the total energy flow in the modes as a function of the wavenumber $\beta$. These results confirm our discussion based on the analysis of the dispersion characteristics.

Increasing the thickness of the cap layer $d_c$ will push the
dispersion of both right- and left-handed modes inside the gap (see
 the curves for $d_c=0.65$cm in Fig.~\rpict{tamm_dispers}), thus
providing better localization of the modes. In the second gap, for
the chosen parameters the LHM mode exists deeper in the gap than
the right-handed mode, allowing for a better wave localization. An
example of the second-gap-mode profile is shown in
Fig.~\rpict{modes}(d). One can see that the second band modes are
generally weaker localized at the interface then the modes from the
first band gap.

\pict[0.9]{fig04}{power_flow}{Total energy flow in RH (dashed) and LH (solid) Tamm modes vs. $\beta$. Cap layer thickness is $d_c = 0.01cm$.}

Finally, in Fig~\rpict{existence_regions} we plot the existence regions for the surface Tamm modes on the parameter plane $(d_c, \beta)$.
Shaded is the area where the surface modes {\em do not exist}.
Increasing the cap layer thickness, $d_c$, we effectively obtain a
dielectric waveguide, one cladding of which is a homogeneous
medium, while the other one is a photonic crystal. In such a case a
typical mode is shown in Fig.~\rpict{modes}(c). The
contours of the corresponding non-existence regions for right-handed
Tamm states are also shown in Fig.~\rpict{existence_regions}. The
regions have qualitatively different shapes. While the lower
non-existence region moves down with increase of $\beta$ preserving
its width, the corresponding right-handed region shifts upwards, decreasing
significantly in width.

\pict{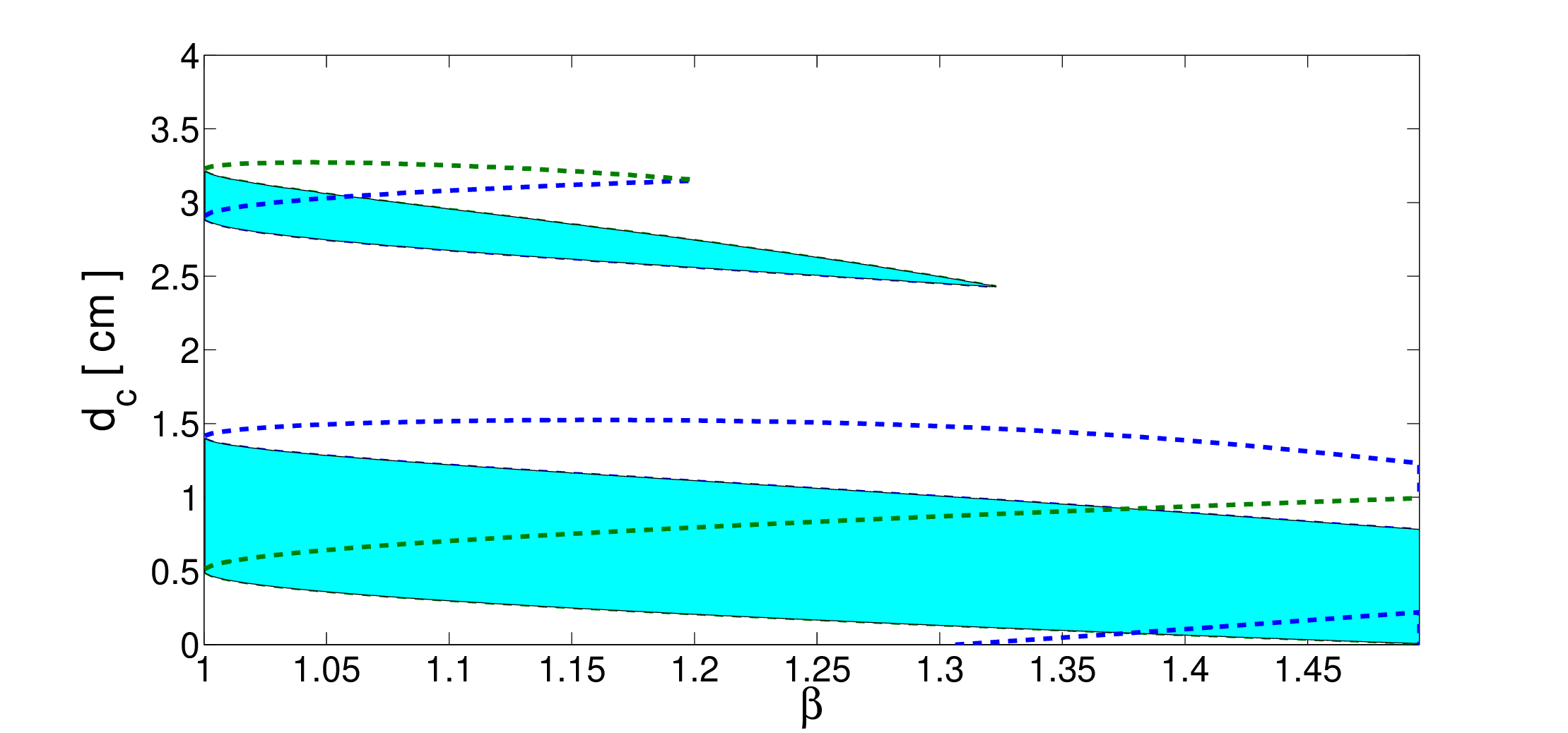}{existence_regions}{Existence regions of the surface Tamm modes; the modes do not exist in the shaded regions. Dashed curves mark the corresponding regions for the right-handed Tamm states.}

In conclusion, we have studied novel types of electromagnetic
surface waves guided by an interface between a left-handed
metamaterial and a one-dimensional photonic crystal. We have shown
that in the presence of a left-handed material the surface Tamm
waves can be either forward or backward while for conventional structures the Tamm states are always forward. We have compared the properties of the backward Tamm states with the case of conventional dielectric structures and analyzed the existence regions for both right- and left-handed Tamm states.
We believe our results will be useful for a deeper understanding of the properties of surface waves in plasmonic and metamaterial systems.

This work was supported by the Australian Research Council and the Azarbaijan University of Tarbiat Moallem.

\end{sloppy}

\begin{thebibliography}{10}
\bibitem{note} Generalization to our results to the case of TM waves is straightforward.
\end{thebibliography}
\end{document}